# Precision measurements of the AC field dependence of the superconducting transition in strontium titanate

Chloe Herrera and Ilya Sochnikov

*Physics Department, University of Connecticut, Storrs, Connecticut, USA 06269*

**Abstract**

Strontium titanate has resurfaced as a material prompting vigorous debate about the origin of its superconductivity in the extremely low carrier concentration regime. Here, we used simultaneous AC susceptibility and transport methods to explore the superconducting phase transition region in this material. We determined that strontium titanate is extremely sensitive to even small AC fields, which also influence the resistive transition; we suggest that extreme vortex sizes and mobilities contribute to this large effect. Our findings will be of importance for accurately determining transition temperature, informing the debate about the pairing mechanism in strontium titanate, for which even millikelvin errors may be critical.

**Main text**

Superconducting pairing in doped strontium titanate (STO) at very low carrier concentrations is well established, yet the mechanism of this superconductivity is a major fundamental open question that has spurred vigorous debate in recent years *(1–12)*. Several recent experimental works have demonstrated the enhancement of superconductivity in STO via compositional *(13–15)* and uniaxial or epitaxial strain techniques *(16, 17)*, and have revealed the potential importance of the ferroelectric quantum phase transition to the mysterious electron pairing *(3, 7, 18–24, 13, 14, 10, 25)*. However, it is not straightforward to compare these recent results with each other, or with classic observations *(18, 19, 26, 27)*, as measurement protocols have varied across investigations. Thus, vigilant attention should be paid when considering the superconducting transition temperature detected with electrical currents of different magnitudes flowing through the sample; different magnetic field amplitudes used in AC susceptibility measurements also merit attention.

In order to elucidate the extent to which these differences matter, we measured transport current and AC field dependence in a single crystal $SrTi_{1-x}Nb_xO_3$ sample, x=0.014 under varying amplitude of the AC field at 200 kHz. Here we report the field-amplitude dependence of this material, and partly review current dependence data that are available elsewhere *(16)*. The size of the reported sample was 5x2x0.5 mm$^3$. We measured the resistance of this sample using a Lakeshore (Ohio, USA) 372 AC resistance bridge, which also served as the temperature controller. The resistance bridge also included a Model 3708 8-channel preamplifier and scanner. When combined with the resistance bridge, this setup yielded an exceptionally low voltage noise floor of 2 nV$_{RMS}$/Hz$^{1/2}$. When supplemented with filtering, a signal level of hundreds of pico-volts could be measured. AC susceptometry was performed using a standard reflection gradiometric design, with miniature pickup coils and a field coil made of copper wire on a G-10 core. AC voltage was detected using SR-830 lockin amplifier (Stanford Research Systems, California, USA).

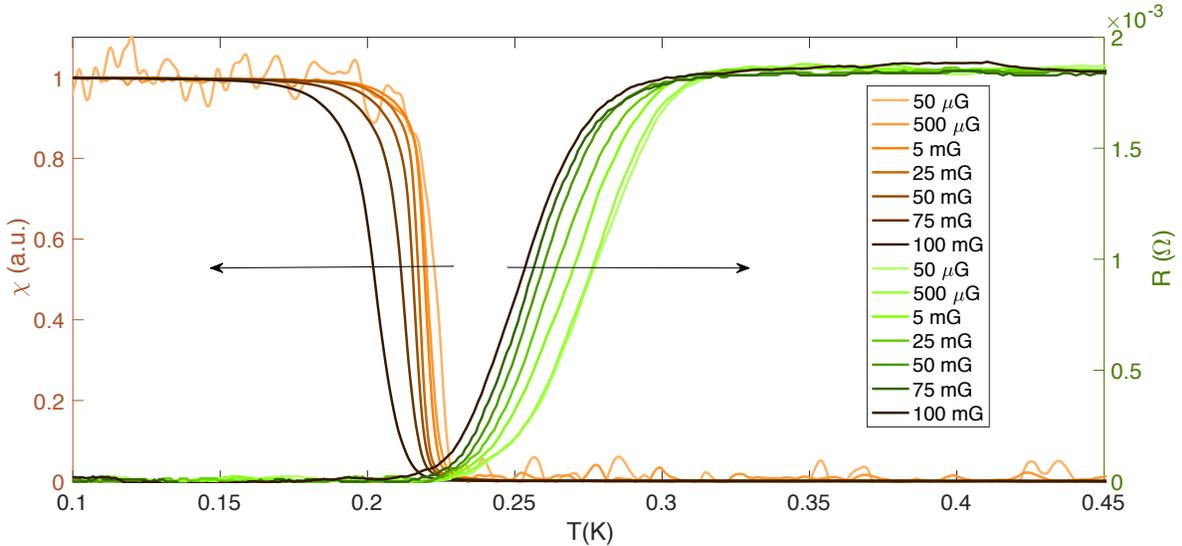

**Figure 1. Simultaneous magnetic and transport measurements of the superconducting phase transition in a single-crystal SrTi$_{1-x}$Nb$_x$O$_3$ sample, x=0.014 under varying amplitude of the AC field at 200 kHz.** The value of 1 on the susceptibility axis (orange, left) corresponds to the full Meissner screening. Resistance (green, right) is measured with 10 μA excitation for all AC fields (key). The transition temperature regions in the susceptibility and resistance measurements are suppressed unusually strongly even under very small AC magnetic fields.

When we varied the field amplitude between 50 μG and 100 mG, our susceptibility and resistance data (**Figure 1**) revealed that the temperature range of the transition is strongly dependent on the AC magnetic field applied to the sample. The observed dependence is unusually strong compared to some other type-II superconductors *(28)*. For fields of 100 mG, the change in the apparent transition temperature was on the order of 10% (Figure 1), which is quite large, given that the reported enhancements of the critical temperatures are on tens of percent levels *(13–17)*. The changes in the apparent transition temperature are reflected in both the resistance features and magnetic features: R(T) and χ(T) shift almost parallel relative to each other and to the temperature axis, toward lower temperatures (Figure 1).

We previously reported *(16)* that curves of resistance versus temperature reveal a substantial contribution from current induced vortex or phase-slips dynamics *(29–31)* close to the transition. This result persists on a similar level regardless of whether the AC field is present.

One of the implications of our investigations is comparisons of published data may be complicated by differences in the excitation currents and the specific criteria used to define the critical temperature. We propose that a variety of normal resistance onsets be used, rather than a single criterion. As long as the excitation current density is known or, even better, constant between investigations, it is sensible to compare transition temperatures as defined using, for example, 2%, 10%, 50%, 90%, or 98% of the normal resistance criteria.

In summary, both transport and AC magnetic susceptibility are often used as basic tools for characterizing the superconducting transition in STO. While the applied excitations can be quite large for many other materials, extra care should be taken with extremely low superfluid density and relatively clean STO – implying very large vortices *(26, 32)* and weak vortex pinning. We demonstrate that dynamic vortex effects are at play in STO, in line with the description of STO as an extreme type-II superconductor *(26, 32)*. While progress is being made in applying theory to better understand the phase transition signatures in STO *(33)*, we still lack a comprehensive theoretical framework for these aspects of superconducting STO. Overall, the physics and experimental aspects discussed here should be relevant to the investigation of other unconventional *(34, 35)* low-concentration superconducting systems and materials.